\documentclass[aps, pra, amsmath, showpacs, preprintnumbers, twocolumn, amssymb]{revtex4-1}
\usepackage{CJK}
\usepackage{braket}
\usepackage{graphicx}
\usepackage{mathptmx, textcomp}
\usepackage[latin1]{inputenc}
\usepackage{tabularx}
\usepackage[usenames]{color}
\usepackage[colorlinks, urlcolor=blue, citecolor=blue, linkcolor=blue]{hyperref}
\begin{document}
\begin{CJK*}{UTF8}{gbsn} 
\title{Observation of the Second Triatomic Resonance in Efimov's Scenario}

\author{Bo Huang (黄博)$^{1}$}
\author{Leonid A. Sidorenkov$^{1,2}$}
\author{Rudolf Grimm$^{1,2}$}
\affiliation{$^1$Institut f\"ur Experimentalphysik, Universit\"at
 Innsbruck, 
 6020 Innsbruck, Austria}
\affiliation{$^2$Institut f\"ur Quantenoptik und Quanteninformation (IQOQI),
 \"Osterreichische Akademie der Wissenschaften, 6020 Innsbruck,
 Austria}

\author{Jeremy M. Hutson}
\affiliation{Joint Quantum Centre (JQC) Durham/Newcastle, Department of Chemistry, Durham University, South Road, Durham, DH1~3LE, United
Kingdom}

\date{\today}
\begin{abstract}
We report the observation of a three-body recombination resonance in an ultracold gas of cesium atoms at a very large negative value of the $s$-wave scattering length. The resonance is identified as the second triatomic Efimov resonance, which corresponds to the situation where the first excited Efimov state appears at the threshold of three free atoms. This observation, together with a finite-temperature analysis and the known first resonance, allows the most accurate demonstration to date of the discrete scaling behavior at the heart of Efimov physics. For the system of three identical bosons, we obtain a scaling factor of $21.0(1.3)$, close to the ideal value of $22.7$.
\end{abstract}
\maketitle
\end{CJK*}

Efimov's prediction of weakly bound three-body states in a
system of three resonantly interacting bosons \cite{Efimov1970ela, Braaten2006uif} is widely known as the paradigm of universal few-body quantum physics. Its bizarre and counterintuitive properties have attracted a great deal of attention. Originally predicted in the context of nuclear systems, Efimov states are now challenging atomic and molecular physics and have strong links to quantum many-body physics \cite{Jensen2011sio}. Experimentally the famous scenario remained elusive until experiments in an ultracold gas of Cs atoms revealed the first signatures of the exotic three-body states \cite{Kraemer2006efe}. A key requirement for the experiments is the precise control of two-body interactions enabled by magnetically tuned Feshbach resonances \cite{Chin2010fri}. With advances in various atomic systems \cite{Zaccanti2009ooa, Pollack2009uit, Gross2009oou, Huckans2009tbr, Ottenstein2008cso, Williams2009efa, Gross2010nsi, Lompe2010ads, Nakajima2010nea, Barontini2009ooh, Roy2013tot, Wild2012mot} and theoretical progress in understanding Efimov states and related states in real systems \cite{Jensen2011sio,wang2013ufb}, the research field of few-body physics with ultracold atoms has emerged.

Three-body recombination resonances \cite{Esry1999rot} are the most prominent signatures of Efimov states \cite{Braaten2006uif, Ferlaino2011eri}. They emerge when an Efimov state
couples to the threshold of free atoms at distinct negative values of the $s$-wave scattering length $a$. The resonance positions $a_-^{(n)}$ are predicted to reflect the discrete scaling law at the heart of Efimov physics, and for the system of three identical bosons follow $a_-^{(n)} = 22.7^n a_-^{(0)}$. Here $n=0$ refers to the Efimov ground state and $n=1, 2, ..$ refer to excited states. The starting point $a_-^{(0)}$ of the infinite series, i.e.\ the position of the ground-state resonance, is commonly referred to as the three-body parameter \cite{Berninger2011uot, Roy2013tot, wang2012oot,Sorensen2012epa,Schmidt2012epb}. 

For an observation of the second Efimov resonance, the requirements are much more demanding than for the first one. Extremely large values of the scattering length near $a_-^{(1)}$ need to be controlled and the relevant energy scale is lower by a factor $22.7^2 \approx 500$, which requires temperatures in the range of a few nK. So far, experimental evidence for an excited-state Efimov resonance has been obtained only in a three-component Fermi gas of $^6$Li \cite{Williams2009efa}, but there the scenario is more complex because of the involvement of three different scattering lengths. Experiments on bosonic $^7$Li have approached suitable conditions for a three-boson system \cite{Pollack2009uit, Dyke2013frc, Rem2013lot} and suggest the possibility of observing the excited-state Efimov resonance \cite{Rem2013lot}.

In this Letter, we report on the observation of the second triatomic resonance in Efimov's original three-boson scenario realized with cesium atoms. Our results confirm the existence of the first excited three-body state and allow the currently most accurate test of the Efimov period. Moreover, our results provide evidence for the existence of the predicted universal $N$-body states that are linked to the excited three-body state.

Two recent advances have prepared the ground for our present investigations. First, we have gained control of very large values of the scattering length (up to a few times $10^5\,a_0$ with $a_0$ being Bohr's radius), which in ultracold Cs gases is achieved by exploiting a broad Feshbach resonance near 800\,G \cite{Lee2007ete, Berninger2011uot}. Precise values for the scattering length as a function of the magnetic field can be obtained from coupled-channel calculations based on the M2012 model potentials of Refs.~\cite{Berninger2011uot,Berninger2013fsf}. Second, Ref.~\cite{Rem2013lot} has provided a model, based on an S-matrix formalism \cite{Efimov1979lep,Braaten2008tbr}, to describe quantitatively the finite-temperature effects on three-body recombination near Efimov resonances. While for the first Efimov resonance experimental conditions can be realized practically in the zero-temperature limit, finite-temperature limitations are unavoidable for the second resonance and therefore must be properly taken into account.

   Our experimental procedure of preparing an ultracold sample of cesium atoms near quantum degeneracy is similar to the one reported in Refs.~\cite{Berninger2011PhD,Berninger2011uot}. 
 In an additional stage, introduced into our setup for the present work, we adiabatically expand the atomic cloud into a very weak trap. The latter is a hybrid with optical confinement by a single infrared laser beam and magnetic confinement provided by the curvature of the magnetic field \cite{SM}.
 The mean oscillation frequency $\bar{\omega}/2 \pi$ of the nearly isotropic trap is about 2.6~Hz. This very low value corresponds to a harmonic oscillator length of $\sim5$~$\mu$m, which is about a factor of five larger than the expected size of the second Efimov state. Our ultracold atomic sample consists of about $N=3\times10^4$ Cs atoms at a temperature of 7~nK and a dimensionless phase-space density of about 0.2.
We probe the atomic cloud by \textit{in-situ} absorption imaging near the zero crossing of the scattering length at 882 G. We obtain the in-trap density profile and the temperature $T$ assuming the gas is thermalized in a harmonic trap. 

To study recombinative decay for different values of the scattering length $a$, we ramp the magnetic field from the final preparation field ($\sim$820 G) \cite{SM} down to a target value (between 818 G and 787 G)~\cite{B_range} within 10 ms. After a variable hold time $t$, between tens of milliseconds and several seconds, we image the remaining atoms. The maximum hold time is chosen to correspond to an atom number decay of about 50\%. In addition to the resulting decay curves $N(t)$ we record the corresponding temperature evolution $T(t)$. Recombinative decay is known to be accompanied by heating~\cite{Weber2003tbr, SM}, which needs to be taken into account when analyzing the results.

For extracting recombination rate coefficients from the observed decay curves, we apply a model that is based on the general differential equation for $\alpha$-body loss in a harmonically trapped thermal gas,
\begin{equation}\label{eq.abody}
\frac{\dot{N}}{N} = - L_{\alpha} \alpha^{-3/2} \left( \frac{N}{V} \right)^{\alpha - 1},
\end{equation}
with the volume $V = \left( 2 \pi k_B T/m \bar{\omega}^2 \right)^{3/2}$. The factor $\alpha^{-3/2}$ arises from the spatial integration of the density-dependent losses.

Since three-body recombination is expected to dominate the decay, we fix $\alpha$ to a value of 3, numerically integrate  Eq.~(\ref{eq.abody}) over time and fit the measured atomic number evolution with $L_3$ and the initial atom number $N_0$ as free parameters. In cases where there are significant contributions from higher-order decay processes, e.g.\ four-body decay, the fitted $L_3$ can be interpreted as an `effective' loss coefficient~\cite{Vonstecher2009sou} that includes all loss processes. Considering a typical temperature change of about 50\% during the decay, a slight complication arises from the fact that $L_3$ itself generally depends on $T$, while our fit assumes constant $L_3$. To a good approximation, however, we can refer a fit value for $L_3$ to a time-averaged temperature $T_{\rm avg}$~\cite{SM} .

\begin{figure}
\includegraphics[width=8cm] {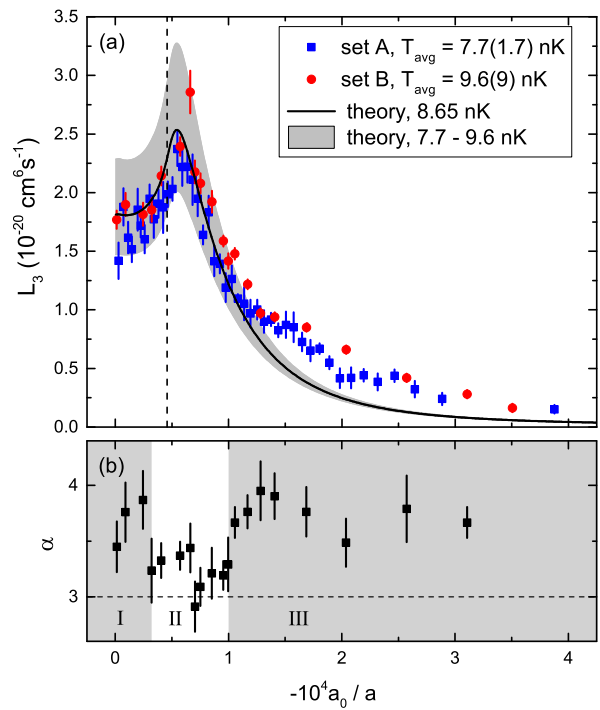}
\caption{Observation of the second triatomic Efimov resonance. In (a), we show the effective three-body loss coefficient $L_{3}$ as a function of the inverse scattering length. Blue squares and red circles are two sets of measurements taken with slightly different trap settings. The error bars include statistical uncertainties from numerical fitting. The black solid line is the theoretical calculation of $L_3$ (based on the parameters of the first Efimov resonance) at 8.65 nK (average temperature of two sets) while the gray-shaded region corresponds to the temperature range between 7.7 and 9.6 nK (see text). The vertical dashed line indicates the position where the resonance would be expected in the zero-temperature limit based on the previously investigated first Efimov resonance~\cite{Berninger2011uot} and Efimov's scaling factor.
In (b), the fitted loss index $\alpha$ shows a larger deviation from a value of 3 in the regions (gray-shaded) away from the second Efimov resonance (white region, where $\alpha<3.5$), indicating  contributions from higher-order recombination processes.}
\label{fig.L3}
\end{figure}

Figure~\ref{fig.L3}(a) shows our main result, the recombination resonance
caused by an excited Efimov state. Here we plot the fit values obtained for
$L_{3}$ as a function of the inverse scattering length $1/a$.
Our sets of measurements (A: blue squares and B: red circles) \cite{SM}
 were taken on different days with similar trap frequencies but
slightly different average temperatures $T_{\rm avg}$ of
7.7(1.7) nK and 9.6(9) nK \cite{SM}. Our results exhibit a loss
peak near $a = -17000a_0$ ($\sim$797 G), which we interpret as a clear
manifestation of the second Efimov resonance. Multiplying
$a_-^{(0)} = -963a_0$~\cite{SM} by Efimov's ideal scaling factor of 22.7
predicts that, in the zero-temperature limit, this feature would occur at
$-21900a_0$ (dashed vertical line in Fig.~\ref{fig.L3}(a)). At finite
temperatures, however, a down-shift towards somewhat lower
values of $|a|$ is expected~\cite{Kraemer2006efe} and may to a large extent explain the observed
position. The finite temperature in our experiment also explains why the
resonance is not as pronounced as the first Efimov resonance observed previously~\cite{Berninger2011uot}.


In order to compare our results with theoretical predictions, we use the
finite-temperature model of Ref.~\cite{Rem2013lot} with the two resonance
parameters, position $a_-^{(0)}= -963 a_0$ and decay parameter $\eta_-^{(0)} =
0.10$ \cite{SM}, independently derived from previous measurements on the first
Efimov resonance. For the temperature we use $T_{\rm avg}$= 8.65 nK, which is
the mean value for the two sets. The agreement between our
present results and the prediction (black
solid line in Fig.~\ref{fig.L3}(a)) is remarkable, and
highlights the discrete scaling behavior of the Efimov scenario.

\begin{figure}[t]
\includegraphics[width=8cm] {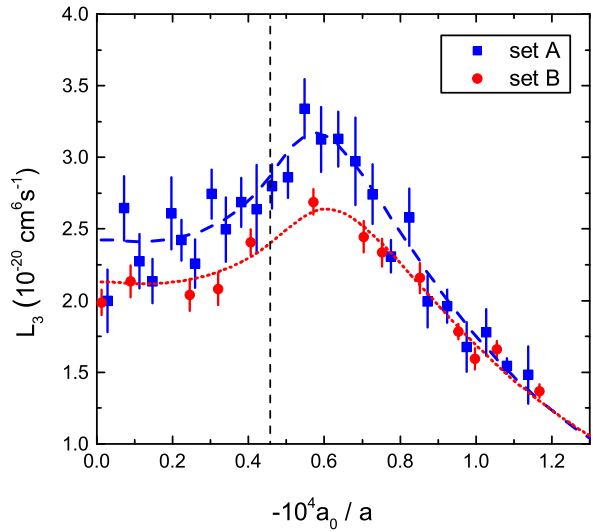}

\caption{Fits to the second Efimov resonance. The two sets of 
points represent the same sets of results as in Fig.~\ref{fig.L3}, but
limited to the resonance region and with one clear outlyer removed.
In addition, the absolute scaling for $L_3$ is changed as we
here use the fit values (see text) for the temperature, 8.7 nK for set A and
10.0 nK for set B, to calculate the volume $V$. The vertical dashed line is the
same as in Fig.~\ref{fig.L3}.
The fits to sets A and B are plotted as blue dashed and red dotted lines,
respectively.} \label{fig.zoom}
\end{figure}

The measurements on the `shoulder' of the resonance ($-10^4a_0/a>1.2$ in
Fig.~\ref{fig.L3}(a)) show a broad increase of the effective $L_{3}$ as
compared to the expectation from the three-body loss theory (black solid line
in Fig.~\ref{fig.L3}(a)). Since similar enhanced loss features were observed previously near the first Efimov resonance~\cite{Ferlaino2009efu,Zenesini2013rfb,Pollack2009uit,Dyke2013frc,Fletcher2013soa}
and were explained by the presence of four- or five-body states
associated with an Efimov state, we attribute this feature to higher-order
decay processes. To check this, we fit set B~\cite{2nd_setB} with
Eq.~(\ref{eq.abody}) as discussed above, while now using $\alpha$ as an
additional free parameter. The fit results for $\alpha$ are shown in
Fig.~\ref{fig.L3}(b). In the region close to the Efimov resonance (white region
II), where we expect dominant three-body behavior, the value of $\alpha$ is
relatively close to 3~\cite{alpha_heat}. On the `shoulder' of the Efimov peak
(gray-shaded region III), a significant increase of $\alpha$, compared to the
resonance region, confirms the existence of higher-order decay processes. It is
interesting to note that the relatively broad shoulder that we observe for the
higher-order features is in contrast to the narrow features observed in
$^7$Li~\cite{Pollack2009uit, Dyke2013frc}. On the other side of the Efimov
resonance (gray-shaded region I), we also observe an enhancement of $\alpha$,
which is likely to be caused by similar higher-order decay features associated
with highly excited $N$-body cluster states.

The temperature uncertainty plays an important role in the
interpretation of our results. The measured values of $L_3$ depend sensitively on the temperature, with a general scaling $\propto T^3$ according to the volume $V$ in Eq.~(\ref{eq.abody}). The theoretical $L_3$ values also depend strongly on the temperature. The gray-shaded
area in Fig.~1(a) demonstrates the variation between 7.7~nK and
9.6~nK, which correspond to $T_{\rm avg}$ for sets A and B,
respectively. It may be seen that the temperature uncertainty results mainly in an amplitude error rather than an error in the peak position.

To analyze the observed resonance in more detail, and especially
to study the possible small deviation of $a_-^{(1)}$ from a predicted value of
22.7$a_-^{(0)}$, we now fit the results in the resonance region ($0<-10^4a_0/a<1.2$ in Fig.~\ref{fig.zoom})
with the finite-temperature model to extract an experimental value for
$a_-^{(1)}$.
Here, because of the large effect of the temperature uncertainty, we use the
temperature $T$ as an additional parameter in the fits. The
results (blue dashed and red dotted lines in Fig.~\ref{fig.zoom} for sets A and
B) are summarized in the upper part of Table~\ref{tab.aminus} and yield a mean
$a_-^{(1)}$ value of $-20270(680) a_0$. 

\begin{table}[t]
\begin{ruledtabular}
\begin{tabular}{lrrrr}
Set  & $T$/nK & $a_{-}^{(1)}$/$a_0$ & $\eta_-^{(1)}$ & $\lambda$\\
\hline
A				& 8.7(2)  & -20790(390) & 0.15(2)	& - \\ 
B				& 10.0(2) & -19740(430) & 0.19(3)	& - \\ 
\hline
A  	 			& 7.7$^*$ & -20580(390) & 0.17(3) 	& 0.52(5) \\
B	 	 		& 9.6$^*$ & -19650(430) & 0.19(3) 	& 0.80(7) \\
\end{tabular}
\end{ruledtabular}
\caption{Fitted parameters for the second Efimov resonance.
The upper part of the table shows the fitting results when temperature $T$ is a free parameter,
 while the lower part corresponds to fixed-temperature fitting with $\lambda$ as a free amplitude scaling factor. The uncertainties indicate $1\sigma$ errors from fitting.
 The symbol $^*$ indicates that the corresponding parameter is kept fixed.
 }
\label{tab.aminus}
\end{table}

%


The fitted results for the temperature, 8.7(2)~nK for set A and 10.0(2)~nK for set B, are somewhat larger than the independently determined temperatures $T_{\rm avg}$, but they are consistent with $T_{\rm avg}$ within the error range. The higher temperatures also imply a rescaling of the measured $L_3$ values because of the temperature dependence of the volume $V$. With these corrections, Fig.~\ref{fig.zoom} shows that the measurements of set A, taken at a lower temperature, now produce larger $L_3$ values than those of set~B.

Uncertainties in $L_3$ might also arise from errors in the atom number calibration,
 resulting from imaging imperfections and
errors in trap frequency measurements. To account for these effects, we follow
an alternative fitting strategy and introduce an additional
parameter $\lambda$ as an amplitude scaling factor for $L_3$ into the
finite-temperature model, while fixing the temperature at the
measured $T_{\rm avg}$. The resulting parameters for each set are
given in the lower part of Table~\ref{tab.aminus}. Remarkably, this
alternative approach gives a mean value of $-20120(630)a_0$ for $a_-^{(1)}$,
which is consistent with the one extracted before. This shows the robustness of
our result for $a_-^{(1)}$. From all the four fits listed in Table~\ref{tab.aminus}, we derive a mean value and a corresponding uncertainty of $-20190(660)a_0$.

A final significant contribution to our error budget for $a_-^{(1)}$ stems from
uncertainty in the M2012 potential model
\cite{Berninger2011uot,Berninger2013fsf} that provides the mapping between the
measured magnetic field $B$ and the scattering length $a$. To quantify this, we
have recalculated the derivatives of all the experimental quantities fitted in
Ref.~\cite{Berninger2013fsf} with respect to the potential parameters, and used
them to obtain fully correlated uncertainties in the calculated scattering
lengths at the magnetic fields $B = 852.90$ G and 795.56 G, corresponding to
the two Efimov loss maxima, using the procedure of
Ref.~\cite{Albritton1976ait}. The resulting scattering lengths and their
$1\sigma$ uncertainties are $a_-^{(0)} = -963(6)\ a_0$ and $a_-^{(1)} =
-20190(1000)\ a_0$. These values accord well with the uncertainty in the
position of the Feshbach resonance pole, which was determined to be 786.8(6) G
in Ref.~\cite{Berninger2013fsf} with a $2\sigma$ uncertainty.

Taking all these uncertainties into account, we get $a_-^{(1)} =
-20190(1200)a_0$ and $a_-^{(0)}=-963(11)a_0$, and we finally obtain
$a_-^{(1)}/a_-^{(0)} = 21.0(1.3)$ for the Efimov period. This result is
consistent with the ideal value of 22.7 within a $1.3\sigma$ uncertainty range.
Theories that take the finite interaction range into account consistently 
predict corrections toward somewhat lower values than 22.7~\cite{DIncao2009tsr,Platter2009rct,Thogersen2008upo}. Ref.~\cite{Schmidt2012epb} 
predicts a value of 17.1 in the limit of strongly entrance-channel-dominated Feshbach resonances. This theoretical value differs by $3\sigma$ 
from our experimental result, but the precise value depends at a $10\%$ level on a form factor that accounts for the range of the coupling between the open and closed channels. 
 Universal van der Waals theory~\cite{Wang2014uvd} applied to our specific Feshbach resonance predicts a value 
that is smaller than the ideal Efimov factor by only 5-$10\%$~\cite{wangjulienne}, which would match our observation.


Additional systematic uncertainties may
slightly influence our experimental determination of the Efimov period.
Model-dependence in the earlier fit to various
interaction-dependent observables in Cs \cite{Berninger2013fsf} may somewhat
affect the mapping $a(B)$ from magnetic field to scattering
length. The  finite-temperature model \cite{Rem2013lot} applied
here, which employs the zero-range approximation, may be influenced by small
finite-range corrections. Moreover, confinement-induced effects may play an
additional role even in the very weak trap \cite{Jonsell2002ubo,
Levinsen2014etu}. While an accurate characterization of these possible
systematic effects will require further effort, we estimate that
our error budget is dominated by the statistical uncertainties.


Previous experiments aimed at determining the Efimov period in $^{39}$K \cite{Zaccanti2009ooa} and $^7$Li \cite{Pollack2009uit, Dyke2013frc} considered recombination minima for $a>0$, from which values of $25(4)$ and $16.0(1.3)$ were extracted, respectively. There the lower recombination minima serving as lower reference points appear at quite small values of the scattering length (typically only at $3$ to $4$ times the van der Waals length $R_{\rm vdW}$ \cite{Chin2010fri}), so that substantial quantitative deviations from Efimov's scenario, which is strictly valid only in the zero-range limit $|a|/R_{\rm vdW} \rightarrow \infty$, may be expected. In our case the lower reference point $a_-^{(0)}$ is at about $-9.5\,R_{\rm vdW}$ (with $R_{\rm vdW}=101 a_0$ for Cs)~\cite{Berninger2011uot,  wang2012oot,Sorensen2012epa,Schmidt2012epb}, which makes the situation more robust. Moreover, at negative scattering length possible effects related to a non-universal behavior of the weakly bound dimer state are avoided~\cite{Zenesini2014vdw}. Another difference between our work and previous determinations of the Efimov period is the character of the Feshbach resonance, which in our case is the most extreme case so far discovered of an entrance-channel-dominated resonance, where the whole interaction can be reduced to an effective single-channel model~\cite{Chin2010fri}. The resonances exploited in $^{39}$K and $^7$Li have intermediate character,  so that the interpretation is less straightforward.

In conclusion, our observation of the second triatomic recombination resonance in an ultracold gas of Cs atoms demonstrates the existence of an excited Efimov state. Together with a previous observation of the first resonance and an analysis based on finite-temperature theory, our results provide an accurate quantitative test of Efimov's scenario of three resonantly interacting bosons. The character of the extremely broad Feshbach resonance that we use for interaction tuning avoids complications from the two-channel nature of the problem and brings the situation in a real atomic system as close as possible to Efimov's original idea. The value of $21.0(1.3)$ that we extract for the Efimov period is very close to the ideal value of $22.7$ and represents the most accurate demonstration so far of the discrete scaling behavior  at the heart of Efimov physics. Our results challenge theory to describe accurately the small deviations that occur in real atomic systems.

New possibilities for Efimov physics beyond the original three-boson scenario are opened up by ultracold mixtures with large mass imbalance \cite{Dincao2006eto}. The $^{133}$Cs-$^6$Li mixture, where the Efimov period is reduced to a value of 4.88, has been identified as a particularly interesting system  \cite{Repp2013ooi, Tung2013umo}. Two very recent preprints \cite{Tung2014oog,Pires2014ooe} report the observation of consecutive Efimov resonances in this system.

We thank D. Petrov for stimulating discussions and for providing the source
code for the finite-temperature model. We thank C. Salomon for important
discussions and comments on the manuscript. We further thank B. Rem, Y. Wang,
P.~S. Julienne, J. Levinsen, R. Schmidt and W. Zwerger for fruitful discussions and M. Berninger, A.
Zenesini, H.-C. N\"agerl, and F. Ferlaino for their important contributions in
an earlier stage of the experiments. We acknowledge support by the Austrian
Science Fund FWF within project P23106 and by EPSRC under grant no.\ EP/I012044/1.

\begin{center}
\textbf{Supplemental Material}
\end{center}
\subsection{Sample Preparation}
\begin{figure}
\includegraphics[width=8cm] {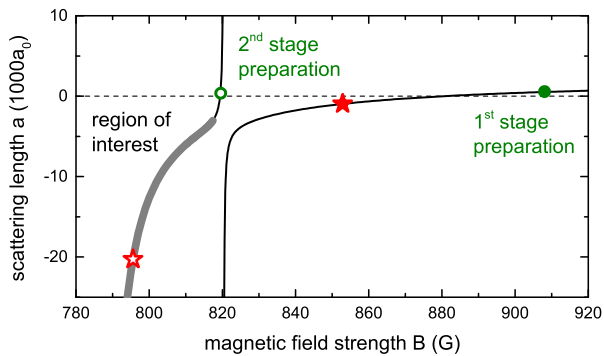}
\caption{Illustration of the tunable region for Cs in the absolute atomic ground state $\Ket{F=3, m_F=3}$ used in the present experiments. The green filled and open circles show the positions for the two stages of sample preparation (see text). The thick gray curve corresponds to the region of interest for observing the second Efimov resonance. The red filled star and the open star indicate the positions of the observed first and second Efimov resonances, respectively.}
\label{fig.Feshbach}
\end{figure}

   Here we discuss the main procedures to create an ultracold near-degenerate sample of Cs atoms  as the starting point for our measurements. We first cool the sample by forced evaporation in a crossed optical-dipole trap at magnetic field near 907 G, where the scattering length $a$ is approximately +500$a_0$ (green filled circle in Fig.~\ref{fig.Feshbach}), and stop slightly before reaching  quantum degeneracy. 
We then adiabatically remove one of the trapping beams and decrease the intensity of the other one to open the trap, thus lowering the temperature of the thermal cloud by adiabatic expansion. 
This transfers the atoms into the resulting hybrid trap, which is formed by one horizontal infrared laser beam and magnetic confinement caused by the curvature of the Feshbach magnetic field. A magnetic gradient field is applied to levitate the atoms in the field of gravity.

For reaching the target magnetic field near 800 G, we have to decrease the magnetic field by about  100~G and cross a narrow Feshbach resonance near 820 G. 
A fast ramp of the magnetic field introduces a drastic change in scattering length and also affects the trapping field. As a result, we observe the excitation of collective oscillations and heating. 
In order to reduce these unwanted effects, we ramp the magnetic field linearly in 20 ms to the positive side of a zero crossing of the scattering length $a$ at 819.4 G (green open circle in Fig.~\ref{fig.Feshbach}, $a\approx +500 a_0$) and stay there for half a second as an intermediate stage of preparation. During this time, the sample thermalizes by collisions and its collective breathing modes damp out. We also adiabatically recompress the optical dipole trap by increasing its power by 50\% to avoid further evaporation during the measurements. Afterwards, we ramp the magnetic field to the target value, which is less than 34 G away, without crossing any Feshbach resonance, thus encountering much weaker effects from heating and excitation of collective modes.

To describe the final trap, we choose a coordinate system as follows: The $z$-axis is collinear with the propagation direction of the laser beam, the $y$-axis is the vertical one, and the $x$-axis is the remaining horizontal one. Using collective sloshing excitations near 819.4 G, we obtain the trap frequencies as $\omega_z/2\pi=$1.34(3)~Hz, $\omega_x/2\pi=$3.97(8)~Hz and $\omega_y/2\pi=$3.36(2)~Hz for measurement set A. The corresponding trap frequencies for set B, which was taken later after small adjustments of the setup, are slightly different: $\omega_z/2\pi=$1.35(4)~Hz, $\omega_x/2\pi=$3.96(8)~Hz and $\omega_y/2\pi=$3.33(7)~Hz. We neglect the slight magnetic field dependence of the trap frequencies, because its effect is smaller than the experimental uncertainties in the magnetic-field range of interest.

\subsection{Density and Temperature Measurement}

Knowledge of the cloud's density profile is essential for obtaining accurate values for the three-body recombination rate coefficient, as well as for thermometry in our experiment. We probe the the Gaussian-shaped thermal atomic cloud by \textit{in-situ} absorption imaging at 881.9~G near a zero crossing of the scattering length $a$. A very short ramp time ($\sim 20$ ms) to this imaging field ensures that the cloud keeps its original spatial distribution. 
The optical axis of the imaging system lies in the horizontal plane, at an angle of $\theta=60^{\circ}$ with respect to the $z$-axis. Therefore, the vertical cloud width obtained from the image is simply the cloud width $w_y$ in the $y$-direction (half $1/e$-width), while the measured horizontal width $w_h$ is related to $w_x$ and $w_z$, the widths in the $x$- and $z$-direction, via $w_h^2={w_z^2\sin^2 \theta+w^2_x\cos^2 \theta}$. We use $w_y$ to calculate the width in the $x$-direction as  $w_x=w_y\omega_y/\omega_x $ and then we extract $w_z$. Since the contribution to $w_h^2$ from the second term is only about 5\%, the extracted $w_z$ only weakly depends on $w_y$ and is mostly defined by $w_h$.


The widths $w_y$ and $w_z$ are used to calculate the cloud temperatures $T_y={m w_y^2 \omega_y^2}/{2 k_B }$ and $T_z={m w_z^2\omega_z^2}/{2 k_B}$. The value of $T_y$ is typically found to be 10\% higher than that of $T_z$, which may be caused by residual collective breathing excitations in the cloud and by the limited resolution of the imaging system. We finally take the mean value $T=(T_y+T_z)/2$ with a corresponding error bar for the further analysis.

\subsection{Fitting of Decay Curves}

\begin{figure}[b]
\includegraphics[width=1\columnwidth] {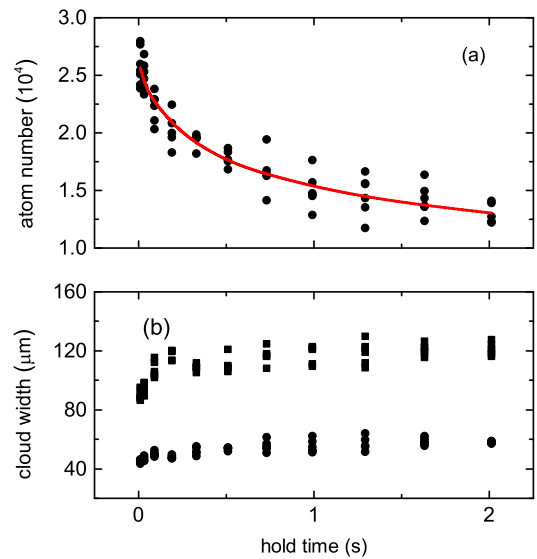}
\caption{Typical data of an atom decay measurement (at 800.87~G or $a=-11800 a_0$) near the second Efimov resonance, taken from set B. 
In (a), we plot measured atom number (black squares) against hold time. In (b), the cloud's widths in horizontal and vertical directions are plotted as black squares and dots, respectively. The red solid curve in (a) is the result of numerical fitting used to extract $L_3$.}
\label{fig.2ndDecay}
\end{figure}

During the decay process, we observe a temperature increase, which results from antievaporation~\cite{Weber2003tbr}, parametric heating in the trap, and heating caused by the damping of the residual collective excitations. In Fig.~\ref{fig.2ndDecay} we show typical data for an atom decay measurement near the second Efimov resonance, taken from measurement set B at 800.57 G.  The substantial decay of the atom number, shown in Fig.~\ref{fig.2ndDecay}(a), is accompanied by an increase of the cloud widths in horizontal ($w_z$) and vertical ($w_y$) directions; see  Fig.~\ref{fig.2ndDecay}(b). 
For measurement sets A and B, the typical increase in width is about 20\% and 30\%, respectively, corresponding to an increase of 45\% and 70\% in temperature.

We numerically solve the differential equation $\dot{N}/N=-0.192\ L_3 (N/V)^2$, where the values for the time-dependent cloud volume $V$ are obtained from the interpolation of measured $V=\pi^{3/2}w_xw_yw_z$ at different hold times. We extract $L_3$ by fitting the calculated $N(t)$ to the experimental data with $L_3$ and the initial atom number $N_0$ being free parameters.


In the fitting procedure, $L_3$ is assumed to be a constant during the decay process, while in reality it changes when the temperature increases. To compare the fitted $L_3$ with the theoretical expectations for a fixed temperature, we introduce for each atom decay measurement a time-averaged temperature $T_{\rm avg}=\left[T(t_1)+2\sum^{n-1}_{i=2}T(t_i)+T(t_n)\right]/2(n-1)$, where the sum is taken over the $n$ different hold times of the individual decay curves. 
In principle, each $L_3$ measurement has its own $T_{\rm avg}$, but within one set (A or B) the variations are small and mostly of statistical nature. Therefore, we characterize each measurement set by its mean value of $T_{\rm avg}$.

\subsection{Details on Measurement Sets A and B}

Between the acquisitions of set~A and B, we slightly adjusted the trap and carried out a routine optimization procedure of the imaging system. Moreover, for each point in set A, the maximum hold time is about 0.5~s and the maximum atom number loss is about 30\% while for set B, the values are 2~s and 50\%.  
Furthermore, in the case of data set~A, atom number and cloud widths are measured at 3 different hold times and repeated for about 10 times. In the case of data set B, the maximum hold time is about 2 s and measurements are done at 11 different hold times and repeated for about 6 times. 

The initial temperatures of set~A and B are 6.1(1.5)~nK and 7.2(5)~nK and the time-averaged temperatures $T_{\rm avg}$ are 7.7(1.7)~nK and 9.6(9)~nK, respectively. The time-averaging of temperature also reduces possible temperature errors caused by residual breathing collective excitation when the data points at different hold times well sample a few oscillation periods.  
Compared to set~A, set~B has a similar sampling rate but a longer sampling time covering more oscillation periods. This makes the measured temperature of set~B somewhat more reliable than that of   set~A.

\begin{figure}[t]
\includegraphics[width=1\columnwidth] {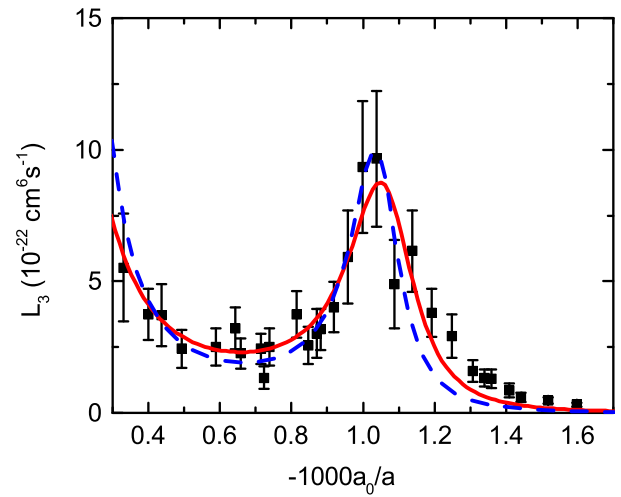}
\caption{Re-fit of the first Efimov resonance. The data from Ref.~\cite{Berninger2011uot} are plotted as black squares. The earlier zero-temperature fit and the new finite-temperature fit are represented by the blue dashed line and the red solid line, respectively.}
\label{fig.refit}
\end{figure}

\subsection{Re-analyzing the First Efimov Resonance}
The $L_3$ data on the first Efimov resonance presented in Ref.~\cite{Berninger2011uot} (black squares in Fig~\ref{fig.refit}) were previously fitted using a \textit{zero-temperature} model with parameters $a_-^{(0)}$ and $\eta_-^{(0)}$ (same as for the model used in present work) and an additional parameter $\lambda$, which is an amplitude scaling factor for $L_3$ accounting for a possible systematic errors in the number density calibration. The result reported in Ref.~\cite{Berninger2011uot} is $a_-^{(0)}=-955(28) a_0$, $\eta_-^{(0)}=0.08(1)$ and $\lambda=0.89(6)$ (black dashed line in Fig~\ref{fig.refit}). In the present work, we fix the temperature to the measured value of 15~nK and refit the data with the \textit{finite-temperature} model~\cite{Rem2013lot} and obtain more precise result as $a_-^{(0)}=-963(9) a_0$, $\eta_-^{(0)}=0.10(1)$ and $\lambda=1.24(6)$ (red line in Fig~\ref{fig.refit}). Here the given errors do not account for uncertainties in the $a(B)$ mapping, as discussed in the main text. The difference in the most important parameter $a_-^{(0)}$ between two fitting approaches is, however, smaller than the 1$\sigma$ uncertainty of the fitting. We conclude that finite-temperature effects have not significantly affected our previous determination of the three-body parameter.


\begin{thebibliography}{10}%
\makeatletter
\providecommand \@ifxundefined [1]{%
 \ifx #1\undefined \expandafter \@firstoftwo
 \else \expandafter \@secondoftwo
\fi
}%
\providecommand \@ifnum [1]{%
 \ifnum #1\expandafter \@firstoftwo
 \else \expandafter \@secondoftwo
\fi
}%
\providecommand \enquote [1]{``#1''}%
\providecommand \bibnamefont  [1]{#1}%
\providecommand \bibfnamefont [1]{#1}%
\providecommand \citenamefont [1]{#1}%
\providecommand\href[0]{\@sanitize\@href}%
\providecommand\@href[1]{\endgroup\@@startlink{#1}\endgroup\@@href}%
\providecommand\@@href[1]{#1\@@endlink}%
\providecommand \@sanitize [0]{\begingroup\catcode`\&12\catcode`\#12\relax}%
\@ifxundefined \pdfoutput {\@firstoftwo}{%
 \@ifnum{\z@=\pdfoutput}{\@firstoftwo}{\@secondoftwo}%
}{%
 \providecommand\@@startlink[1]{\leavevmode}%
 \providecommand\@@endlink[0]{}%
}{%
 \providecommand\@@startlink[1]{%
  \leavevmode
  \pdfstartlink
   attr{/Border[0 0 1 ]/H/I/C[0 1 1]}%
   user{/Subtype/Link/A<</Type/Action/S/URI/URI(#1)>>}%
  \relax
 }%
 \providecommand\@@endlink[0]{\pdfendlink}%
}%
\providecommand \url  [0]{\begingroup\@sanitize \@url }%
\providecommand \@url [1]{\endgroup\@href {#1}{\urlprefix}}%
\providecommand \urlprefix [0]{URL }%
\providecommand \Eprint[0]{\href }%
\@ifxundefined \urlstyle {%
  \providecommand \doi [1]{doi:\discretionary{}{}{}#1}%
}{%
  \providecommand \doi [0]{doi:\discretionary{}{}{}\begingroup
  \urlstyle{rm}\Url }%
}%
\providecommand \doibase [0]{http://dx.doi.org/}%
\providecommand \Doi[1]{\href{\doibase#1}}%
\providecommand \bibAnnote [3]{%
  \BibitemShut{#1}%
  \begin{quotation}\noindent
    \textsc{Key:}\ #2\\\textsc{Annotation:}\ #3%
  \end{quotation}%
}%
\providecommand \bibAnnoteFile [2]{%
  \IfFileExists{#2}{\bibAnnote {#1} {#2} {\input{#2}}}{}%
}%
\providecommand \typeout [0]{\immediate \write \m@ne }%
\providecommand \selectlanguage [0]{\@gobble}%
\providecommand \bibinfo [0]{\@secondoftwo}%
\providecommand \bibfield [0]{\@secondoftwo}%
\providecommand \translation [1]{[#1]}%
\providecommand \BibitemOpen[0]{}%
\providecommand \bibitemStop [0]{}%
\providecommand \bibitemNoStop [0]{.\EOS\space}%
\providecommand \EOS [0]{\spacefactor3000\relax}%
\providecommand \BibitemShut [1]{\csname bibitem#1\endcsname}%
\bibitem{Efimov1970ela}%
  \BibitemOpen
  \bibfield{author}{%
  \bibinfo {author} {\bibfnamefont{V.}~\bibnamefont{Efimov}},\ }%
  \bibfield{journal}{%
  \bibinfo {journal} {Phys. Lett. B}\ }%
  \textbf{\bibinfo {volume} {33}},\ \bibinfo {pages} {563} (\bibinfo {year}
  {1970})%
  \bibAnnoteFile{NoStop}{Efimov1970ela}%
\bibitem{Braaten2006uif}%
  \BibitemOpen
  \bibfield{author}{%
  \bibinfo {author} {\bibfnamefont{E.}~\bibnamefont{Braaten}}\ and\ \bibinfo
  {author} {\bibfnamefont{H.-W.}\ \bibnamefont{Hammer}},\ }%
  \bibfield{journal}{%
  \Doi{10.1016/j.physrep.2006.03.001}{\bibinfo {journal} {Phys. Rep.}}\ }%
  \textbf{\bibinfo {volume} {428}},\ \bibinfo {pages} {259} (\bibinfo {year}
  {2006})%
  \bibAnnoteFile{NoStop}{Braaten2006uif}%
\bibitem{Jensen2011sio}%
  \BibitemOpen
  \bibfield{author}{%
  \bibinfo {author} {\bibnamefont{{A. S. Jensen (editor)}}},\ }%
  \bibinfo {howpublished} {Few-Body Syst. {\bf 51}, 77-269} (\bibinfo {year}
  {2011})%
  \bibAnnoteFile{NoStop}{Jensen2011sio}%
\bibitem{Kraemer2006efe}%
  \BibitemOpen
  \bibfield{author}{%
  \bibinfo {author} {\bibfnamefont{T.}~\bibnamefont{Kraemer}}, \bibinfo
  {author} {\bibfnamefont{M.}~\bibnamefont{Mark}}, \bibinfo {author}
  {\bibfnamefont{P.}~\bibnamefont{Waldburger}}, \bibinfo {author}
  {\bibfnamefont{J.~G.}\ \bibnamefont{Danzl}}, \bibinfo {author}
  {\bibfnamefont{C.}~\bibnamefont{Chin}}, \bibinfo {author}
  {\bibfnamefont{B.}~\bibnamefont{Engeser}}, \bibinfo {author}
  {\bibfnamefont{A.~D.}\ \bibnamefont{Lange}}, \bibinfo {author}
  {\bibfnamefont{K.}~\bibnamefont{Pilch}}, \bibinfo {author}
  {\bibfnamefont{A.}~\bibnamefont{Jaakkola}}, \bibinfo {author}
  {\bibfnamefont{H.-C.}\ \bibnamefont{N\"agerl}},\ and\ \bibinfo {author}
  {\bibfnamefont{R.}~\bibnamefont{Grimm}},\ }%
  \bibfield{journal}{%
  \Doi{10.1038/nature04626}{\bibinfo {journal} {Nature}}\ }%
  \textbf{\bibinfo {volume} {440}},\ \bibinfo {pages} {315} (\bibinfo {year}
  {2006})%
  \bibAnnoteFile{NoStop}{Kraemer2006efe}%
\bibitem{Chin2010fri}%
  \BibitemOpen
  \bibfield{author}{%
  \bibinfo {author} {\bibfnamefont{C.}~\bibnamefont{Chin}}, \bibinfo {author}
  {\bibfnamefont{R.}~\bibnamefont{Grimm}}, \bibinfo {author}
  {\bibfnamefont{P.~S.}\ \bibnamefont{Julienne}},\ and\ \bibinfo {author}
  {\bibfnamefont{E.}~\bibnamefont{Tiesinga}},\ }%
  \bibfield{journal}{%
  \Doi{10.1103/RevModPhys.82.1225}{\bibinfo {journal} {Rev. Mod. Phys.}}\ }%
  \textbf{\bibinfo {volume} {82}},\ \bibinfo {pages} {1225} (\bibinfo {year}
  {2010})%
  \bibAnnoteFile{NoStop}{Chin2010fri}%
\bibitem{Zaccanti2009ooa}%
  \BibitemOpen
  \bibfield{author}{%
  \bibinfo {author} {\bibfnamefont{M.}~\bibnamefont{Zaccanti}}, \bibinfo
  {author} {\bibfnamefont{B.}~\bibnamefont{Deissler}}, \bibinfo {author}
  {\bibfnamefont{C.}~\bibnamefont{D'Errico}}, \bibinfo {author}
  {\bibfnamefont{M.}~\bibnamefont{Fattori}}, \bibinfo {author}
  {\bibfnamefont{M.}~\bibnamefont{Jona-Lasinio}}, \bibinfo {author}
  {\bibfnamefont{S.}~\bibnamefont{M\"uller}}, \bibinfo {author}
  {\bibfnamefont{G.}~\bibnamefont{Roati}}, \bibinfo {author}
  {\bibfnamefont{M.}~\bibnamefont{Inguscio}},\ and\ \bibinfo {author}
  {\bibfnamefont{G.}~\bibnamefont{Modugno}},\ }%
  \bibfield{journal}{%
  \Doi{10.1038/nphys1334}{\bibinfo {journal} {Nature Phys.}}\ }%
  \textbf{\bibinfo {volume} {5}},\ \bibinfo {pages} {586} (\bibinfo {year}
  {2009})%
  \bibAnnoteFile{NoStop}{Zaccanti2009ooa}%
\bibitem{Pollack2009uit}%
  \BibitemOpen
  \bibfield{author}{%
  \bibinfo {author} {\bibfnamefont{S.~E.}\ \bibnamefont{Pollack}}, \bibinfo
  {author} {\bibfnamefont{D.}~\bibnamefont{Dries}},\ and\ \bibinfo {author}
  {\bibfnamefont{R.~G.}\ \bibnamefont{Hulet}},\ }%
  \bibfield{journal}{%
  \Doi{10.1126/science.1182840}{\bibinfo {journal} {Science}}\ }%
  \textbf{\bibinfo {volume} {326}},\ \bibinfo {pages} {1683} (\bibinfo {year}
  {2009})%
  \bibAnnoteFile{NoStop}{Pollack2009uit}%
\bibitem{Gross2009oou}%
  \BibitemOpen
  \bibfield{author}{%
  \bibinfo {author} {\bibfnamefont{N.}~\bibnamefont{Gross}}, \bibinfo {author}
  {\bibfnamefont{Z.}~\bibnamefont{Shotan}}, \bibinfo {author}
  {\bibfnamefont{S.}~\bibnamefont{Kokkelmans}},\ and\ \bibinfo {author}
  {\bibfnamefont{L.}~\bibnamefont{Khaykovich}},\ }%
  \bibfield{journal}{%
  \Doi{10.1103/PhysRevLett.103.163202}{\bibinfo {journal} {Phys. Rev. Lett.}}\
  }%
  \textbf{\bibinfo {volume} {103}},\ \bibinfo {pages} {163202} (\bibinfo {year}
  {2009})%
  \bibAnnoteFile{NoStop}{Gross2009oou}%
\bibitem{Huckans2009tbr}%
  \BibitemOpen
  \bibfield{author}{%
  \bibinfo {author} {\bibfnamefont{J.~H.}\ \bibnamefont{Huckans}}, \bibinfo
  {author} {\bibfnamefont{J.~R.}\ \bibnamefont{Williams}}, \bibinfo {author}
  {\bibfnamefont{E.~L.}\ \bibnamefont{Hazlett}}, \bibinfo {author}
  {\bibfnamefont{R.~W.}\ \bibnamefont{Stites}},\ and\ \bibinfo {author}
  {\bibfnamefont{K.~M.}\ \bibnamefont{O'Hara}},\ }%
  \bibfield{journal}{%
  \Doi{10.1103/PhysRevLett.102.165302}{\bibinfo {journal} {Phys. Rev. Lett.}}\
  }%
  \textbf{\bibinfo {volume} {102}},\ \bibinfo {eid} {165302} (\bibinfo {year}
  {2009})%
  \bibAnnoteFile{NoStop}{Huckans2009tbr}%
\bibitem{Ottenstein2008cso}%
  \BibitemOpen
  \bibfield{author}{%
  \bibinfo {author} {\bibfnamefont{T.~B.}\ \bibnamefont{Ottenstein}}, \bibinfo
  {author} {\bibfnamefont{T.}~\bibnamefont{Lompe}}, \bibinfo {author}
  {\bibfnamefont{M.}~\bibnamefont{Kohnen}}, \bibinfo {author}
  {\bibfnamefont{A.~N.}\ \bibnamefont{Wenz}},\ and\ \bibinfo {author}
  {\bibfnamefont{S.}~\bibnamefont{Jochim}},\ }%
  \bibfield{journal}{%
  \Doi{10.1103/PhysRevLett.101.203202}{\bibinfo {journal} {Phys. Rev. Lett.}}\
  }%
  \textbf{\bibinfo {volume} {101}},\ \bibinfo {eid} {203202} (\bibinfo {year}
  {2008})%
  \bibAnnoteFile{NoStop}{Ottenstein2008cso}%
\bibitem{Williams2009efa}%
  \BibitemOpen
  \bibfield{author}{%
  \bibinfo {author} {\bibfnamefont{J.~R.}\ \bibnamefont{Williams}}, \bibinfo
  {author} {\bibfnamefont{E.~L.}\ \bibnamefont{Hazlett}}, \bibinfo {author}
  {\bibfnamefont{J.~H.}\ \bibnamefont{Huckans}}, \bibinfo {author}
  {\bibfnamefont{R.~W.}\ \bibnamefont{Stites}}, \bibinfo {author}
  {\bibfnamefont{Y.}~\bibnamefont{Zhang}},\ and\ \bibinfo {author}
  {\bibfnamefont{K.~M.}\ \bibnamefont{O'Hara}},\ }%
  \bibfield{journal}{%
  \Doi{10.1103/PhysRevLett.103.130404}{\bibinfo {journal} {Phys. Rev. Lett.}}\
  }%
  \textbf{\bibinfo {volume} {103}},\ \bibinfo {pages} {130404} (\bibinfo {year}
  {2009})%
  \bibAnnoteFile{NoStop}{Williams2009efa}%
\bibitem{Gross2010nsi}%
  \BibitemOpen
  \bibfield{author}{%
  \bibinfo {author} {\bibfnamefont{N.}~\bibnamefont{Gross}}, \bibinfo {author}
  {\bibfnamefont{Z.}~\bibnamefont{Shotan}}, \bibinfo {author}
  {\bibfnamefont{S.}~\bibnamefont{Kokkelmans}},\ and\ \bibinfo {author}
  {\bibfnamefont{L.}~\bibnamefont{Khaykovich}},\ }%
  \bibfield{journal}{%
  \Doi{10.1103/PhysRevLett.105.103203}{\bibinfo {journal} {Phys. Rev. Lett}}\
  }%
  \textbf{\bibinfo {volume} {105}},\ \bibinfo {pages} {103203} (\bibinfo {year}
  {2010})%
  \bibAnnoteFile{NoStop}{Gross2010nsi}%
\bibitem{Lompe2010ads}%
  \BibitemOpen
  \bibfield{author}{%
  \bibinfo {author} {\bibfnamefont{T.}~\bibnamefont{Lompe}}, \bibinfo {author}
  {\bibfnamefont{T.~B.}\ \bibnamefont{Ottenstein}}, \bibinfo {author}
  {\bibfnamefont{F.}~\bibnamefont{Serwane}}, \bibinfo {author}
  {\bibfnamefont{K.}~\bibnamefont{Viering}}, \bibinfo {author}
  {\bibfnamefont{A.~N.}\ \bibnamefont{Wenz}}, \bibinfo {author}
  {\bibfnamefont{G.}~\bibnamefont{Z\"{u}rn}},\ and\ \bibinfo {author}
  {\bibfnamefont{S.}~\bibnamefont{Jochim}},\ }%
  \bibfield{journal}{%
  \Doi{10.1103/PhysRevLett.105.103201}{\bibinfo {journal} {Phys. Rev. Lett}}\
  }%
  \textbf{\bibinfo {volume} {105}},\ \bibinfo {pages} {103201} (\bibinfo {year}
  {2010})%
  \bibAnnoteFile{NoStop}{Lompe2010ads}%
\bibitem{Nakajima2010nea}%
  \BibitemOpen
  \bibfield{author}{%
  \bibinfo {author} {\bibfnamefont{S.}~\bibnamefont{Nakajima}}, \bibinfo
  {author} {\bibfnamefont{M.}~\bibnamefont{Horikoshi}}, \bibinfo {author}
  {\bibfnamefont{T.}~\bibnamefont{Mukaiyama}}, \bibinfo {author}
  {\bibfnamefont{P.}~\bibnamefont{Naidon}},\ and\ \bibinfo {author}
  {\bibfnamefont{M.}~\bibnamefont{Ueda}},\ }%
  \bibfield{journal}{%
  \Doi{10.1103/PhysRevLett.105.023201}{\bibinfo {journal} {Phys. Rev. Lett}}\
  }%
  \textbf{\bibinfo {volume} {105}},\ \bibinfo {pages} {023201} (\bibinfo {year}
  {2010})%
  \bibAnnoteFile{NoStop}{Nakajima2010nea}%
\bibitem{Barontini2009ooh}%
  \BibitemOpen
  \bibfield{author}{%
  \bibinfo {author} {\bibfnamefont{G.}~\bibnamefont{Barontini}}, \bibinfo
  {author} {\bibfnamefont{C.}~\bibnamefont{Weber}}, \bibinfo {author}
  {\bibfnamefont{F.}~\bibnamefont{Rabatti}}, \bibinfo {author}
  {\bibfnamefont{J.}~\bibnamefont{Catani}}, \bibinfo {author}
  {\bibfnamefont{G.}~\bibnamefont{Thalhammer}}, \bibinfo {author}
  {\bibfnamefont{M.}~\bibnamefont{Inguscio}},\ and\ \bibinfo {author}
  {\bibfnamefont{F.}~\bibnamefont{Minardi}},\ }%
  \bibfield{journal}{%
  \Doi{10.1103/PhysRevLett.103.043201}{\bibinfo {journal} {Phys. Rev. Lett.}}\
  }%
  \textbf{\bibinfo {volume} {103}},\ \bibinfo {pages} {043201} (\bibinfo {year}
  {2009})%
  \bibAnnoteFile{NoStop}{Barontini2009ooh}%
\bibitem{Roy2013tot}%
  \BibitemOpen
  \bibfield{author}{%
  \bibinfo {author} {\bibfnamefont{S.}~\bibnamefont{Roy}}, \bibinfo {author}
  {\bibfnamefont{M.}~\bibnamefont{Landini}}, \bibinfo {author}
  {\bibfnamefont{A.}~\bibnamefont{Trenkwalder}}, \bibinfo {author}
  {\bibfnamefont{G.}~\bibnamefont{Semeghini}}, \bibinfo {author}
  {\bibfnamefont{G.}~\bibnamefont{Spagnolli}}, \bibinfo {author}
  {\bibfnamefont{A.}~\bibnamefont{Simoni}}, \bibinfo {author}
  {\bibfnamefont{M.}~\bibnamefont{Fattori}}, \bibinfo {author}
  {\bibfnamefont{M.}~\bibnamefont{Inguscio}},\ and\ \bibinfo {author}
  {\bibfnamefont{G.}~\bibnamefont{Modugno}},\ }%
  \bibfield{journal}{%
  \Doi{10.1103/PhysRevLett.111.053202}{\bibinfo {journal} {Phys. Rev. Lett.}}\
  }%
  \textbf{\bibinfo {volume} {111}},\ \bibinfo {pages} {053202} (\bibinfo {year}
  {2013})%
  \bibAnnoteFile{NoStop}{Roy2013tot}%
\bibitem{Wild2012mot}%
  \BibitemOpen
  \bibfield{author}{%
  \bibinfo {author} {\bibfnamefont{R.~J.}\ \bibnamefont{Wild}}, \bibinfo
  {author} {\bibfnamefont{P.}~\bibnamefont{Makotyn}}, \bibinfo {author}
  {\bibfnamefont{J.~M.}\ \bibnamefont{Pino}}, \bibinfo {author}
  {\bibfnamefont{E.~A.}\ \bibnamefont{Cornell}},\ and\ \bibinfo {author}
  {\bibfnamefont{D.~S.}\ \bibnamefont{Jin}},\ }%
  \bibfield{journal}{%
  \Doi{10.1103/PhysRevLett.108.145305}{\bibinfo {journal} {Phys. Rev. Lett.}}\
  }%
  \textbf{\bibinfo {volume} {108}},\ \bibinfo {pages} {145305} (\bibinfo
  {month} {Apr}\ \bibinfo {year} {2012})%
  \bibAnnoteFile{NoStop}{Wild2012mot}%
\bibitem{wang2013ufb}%
  \BibitemOpen
  \bibfield{author}{%
  \bibinfo {author} {\bibfnamefont{Y.}~\bibnamefont{Wang}}, \bibinfo {author}
  {\bibfnamefont{J.~P.}\ \bibnamefont{D'Incao}},\ and\ \bibinfo {author}
  {\bibfnamefont{B.~D.}\ \bibnamefont{Esry}},\ }%
  \bibfield{journal}{%
  \Doi{http://dx.doi.org/10.1016/B978-0-12-408090-4.00001-3}{\bibinfo {journal}
  {Adv. At. Mol. Opt. Phys.}}\ }%
  \textbf{\bibinfo {volume} {62}},\ \bibinfo {pages} {1} (\bibinfo {year}
  {2013})%
  \bibAnnoteFile{NoStop}{wang2013ufb}%
\bibitem{Esry1999rot}%
  \BibitemOpen
  \bibfield{author}{%
  \bibinfo {author} {\bibfnamefont{B.~D.}\ \bibnamefont{Esry}}, \bibinfo
  {author} {\bibfnamefont{C.~H.}\ \bibnamefont{Greene}},\ and\ \bibinfo
  {author} {\bibfnamefont{J.~P.}\ \bibnamefont{Burke}},\ }%
  \bibfield{journal}{%
  \Doi{10.1103/PhysRevLett.83.1751}{\bibinfo {journal} {Phys. Rev. Lett.}}\ }%
  \textbf{\bibinfo {volume} {83}},\ \bibinfo {pages} {1751} (\bibinfo {year}
  {1999})%
  \bibAnnoteFile{NoStop}{Esry1999rot}%
\bibitem{Ferlaino2011eri}%
  \BibitemOpen
  \bibfield{author}{%
  \bibinfo {author} {\bibfnamefont{F.}~\bibnamefont{Ferlaino}}, \bibinfo
  {author} {\bibfnamefont{A.}~\bibnamefont{Zenesini}}, \bibinfo {author}
  {\bibfnamefont{M.}~\bibnamefont{Berninger}}, \bibinfo {author}
  {\bibfnamefont{B.}~\bibnamefont{Huang}}, \bibinfo {author}
  {\bibfnamefont{H.-C.}\ \bibnamefont{N\"{a}gerl}},\ and\ \bibinfo {author}
  {\bibfnamefont{R.}~\bibnamefont{Grimm}},\ }%
  \bibfield{journal}{%
  \Doi{10.1007/s00601-011-0260-7}{\bibinfo {journal} {Few-Body Syst.}}\ }%
  \textbf{\bibinfo {volume} {51}},\ \bibinfo {pages} {113} (\bibinfo {year}
  {2011})%
  \bibAnnoteFile{NoStop}{Ferlaino2011eri}%
\bibitem{Berninger2011uot}%
  \BibitemOpen
  \bibfield{author}{%
  \bibinfo {author} {\bibfnamefont{M.}~\bibnamefont{Berninger}}, \bibinfo
  {author} {\bibfnamefont{A.}~\bibnamefont{Zenesini}}, \bibinfo {author}
  {\bibfnamefont{B.}~\bibnamefont{Huang}}, \bibinfo {author}
  {\bibfnamefont{W.}~\bibnamefont{Harm}}, \bibinfo {author}
  {\bibfnamefont{H.-C.}\ \bibnamefont{N\"{a}gerl}}, \bibinfo {author}
  {\bibfnamefont{F.}~\bibnamefont{Ferlaino}}, \bibinfo {author}
  {\bibfnamefont{R.}~\bibnamefont{Grimm}}, \bibinfo {author}
  {\bibfnamefont{P.~S.}\ \bibnamefont{Julienne}},\ and\ \bibinfo {author}
  {\bibfnamefont{J.~M.}\ \bibnamefont{Hutson}},\ }%
  \bibfield{journal}{%
  \Doi{10.1103/PhysRevLett.107.120401}{\bibinfo {journal} {Phys. Rev. Lett.}}\
  }%
  \textbf{\bibinfo {volume} {107}},\ \bibinfo {pages} {120401} (\bibinfo {year}
  {2011})%
  \bibAnnoteFile{NoStop}{Berninger2011uot}%
\bibitem{wang2012oot}%
  \BibitemOpen
  \bibfield{author}{%
  \bibinfo {author} {\bibfnamefont{J.}~\bibnamefont{Wang}}, \bibinfo {author}
  {\bibfnamefont{J.~P.}\ \bibnamefont{D'Incao}}, \bibinfo {author}
  {\bibfnamefont{B.~D.}\ \bibnamefont{Esry}},\ and\ \bibinfo {author}
  {\bibfnamefont{C.~H.}\ \bibnamefont{Greene}},\ }%
  \bibfield{journal}{%
  \Doi{10.1103/PhysRevLett.108.263001}{\bibinfo {journal} {Phys. Rev. Lett.}}\
  }%
  \textbf{\bibinfo {volume} {108}},\ \bibinfo {pages} {263001} (\bibinfo
  {month} {Jun}\ \bibinfo {year} {2012})%
  \bibAnnoteFile{NoStop}{wang2012oot}%
\bibitem{Sorensen2012epa}%
  \BibitemOpen
  \bibfield{author}{%
  \bibinfo {author} {\bibfnamefont{P.~K.}\ \bibnamefont{S\o{}rensen}}, \bibinfo
  {author} {\bibfnamefont{D.~V.}\ \bibnamefont{Fedorov}}, \bibinfo {author}
  {\bibfnamefont{A.~S.}\ \bibnamefont{Jensen}},\ and\ \bibinfo {author}
  {\bibfnamefont{N.~T.}\ \bibnamefont{Zinner}},\ }%
  \bibfield{journal}{%
  \Doi{10.1103/PhysRevA.86.052516}{\bibinfo {journal} {Phys. Rev. A}}\ }%
  \textbf{\bibinfo {volume} {86}},\ \bibinfo {pages} {052516} (\bibinfo {month}
  {Nov}\ \bibinfo {year} {2012})%
  \bibAnnoteFile{NoStop}{Sorensen2012epa}%
\bibitem{Schmidt2012epb}%
  \BibitemOpen
  \bibfield{author}{%
  \bibinfo {author} {\bibfnamefont{R.}~\bibnamefont{Schmidt}}, \bibinfo
  {author} {\bibfnamefont{S.}~\bibnamefont{Rath}},\ and\ \bibinfo {author}
  {\bibfnamefont{W.}~\bibnamefont{Zwerger}},\ }%
  \bibfield{journal}{%
  \Doi{10.1140/epjb/e2012-30841-3}{\bibinfo {journal} {Eur. Phys. J. B}}\ }%
  \textbf{\bibinfo {volume} {85}},\ \bibinfo {pages} {386} (\bibinfo {year}
  {2012})%
  \bibAnnoteFile{NoStop}{Schmidt2012epb}%
\bibitem{Dyke2013frc}%
  \BibitemOpen
  \bibfield{author}{%
  \bibinfo {author} {\bibfnamefont{P.}~\bibnamefont{Dyke}}, \bibinfo {author}
  {\bibfnamefont{S.~E.}\ \bibnamefont{Pollack}},\ and\ \bibinfo {author}
  {\bibfnamefont{R.~G.}\ \bibnamefont{Hulet}},\ }%
  \bibfield{journal}{%
  \Doi{10.1103/PhysRevA.88.023625}{\bibinfo {journal} {Phys. Rev. A}}\ }%
  \textbf{\bibinfo {volume} {88}},\ \bibinfo {pages} {023625} (\bibinfo {month}
  {Aug}\ \bibinfo {year} {2013})%
  \bibAnnoteFile{NoStop}{Dyke2013frc}%
\bibitem{Rem2013lot}%
  \BibitemOpen
  \bibfield{author}{%
  \bibinfo {author} {\bibfnamefont{B.~S.}\ \bibnamefont{Rem}}, \bibinfo
  {author} {\bibfnamefont{A.~T.}\ \bibnamefont{Grier}}, \bibinfo {author}
  {\bibfnamefont{I.}~\bibnamefont{Ferrier-Barbut}}, \bibinfo {author}
  {\bibfnamefont{U.}~\bibnamefont{Eismann}}, \bibinfo {author}
  {\bibfnamefont{T.}~\bibnamefont{Langen}}, \bibinfo {author}
  {\bibfnamefont{N.}~\bibnamefont{Navon}}, \bibinfo {author}
  {\bibfnamefont{L.}~\bibnamefont{Khaykovich}}, \bibinfo {author}
  {\bibfnamefont{F.}~\bibnamefont{Werner}}, \bibinfo {author}
  {\bibfnamefont{D.~S.}\ \bibnamefont{Petrov}}, \bibinfo {author}
  {\bibfnamefont{F.}~\bibnamefont{Chevy}},\ and\ \bibinfo {author}
  {\bibfnamefont{C.}~\bibnamefont{Salomon}},\ }%
  \bibfield{journal}{%
  \Doi{10.1103/PhysRevLett.110.163202}{\bibinfo {journal} {Phys. Rev. Lett.}}\
  }%
  \textbf{\bibinfo {volume} {110}},\ \bibinfo {pages} {163202} (\bibinfo
  {month} {Apr}\ \bibinfo {year} {2013})%
  \bibAnnoteFile{NoStop}{Rem2013lot}%
\bibitem{Lee2007ete}%
  \BibitemOpen
  \bibfield{author}{%
  \bibinfo {author} {\bibfnamefont{M.~D.}\ \bibnamefont{Lee}}, \bibinfo
  {author} {\bibfnamefont{T.}~\bibnamefont{K\"{o}hler}},\ and\ \bibinfo
  {author} {\bibfnamefont{P.~S.}\ \bibnamefont{Julienne}},\ }%
  \bibfield{journal}{%
  \Doi{10.1103/PhysRevA.76.012720}{\bibinfo {journal} {Phys. Rev. A}}\ }%
  \textbf{\bibinfo {volume} {76}},\ \bibinfo {eid} {012720} (\bibinfo {year}
  {2007})%
  \bibAnnoteFile{NoStop}{Lee2007ete}%
\bibitem{Berninger2013fsf}%
  \BibitemOpen
  \bibfield{author}{%
  \bibinfo {author} {\bibfnamefont{M.}~\bibnamefont{Berninger}}, \bibinfo
  {author} {\bibfnamefont{A.}~\bibnamefont{Zenesini}}, \bibinfo {author}
  {\bibfnamefont{B.}~\bibnamefont{Huang}}, \bibinfo {author}
  {\bibfnamefont{W.}~\bibnamefont{Harm}}, \bibinfo {author}
  {\bibfnamefont{H.-C.}\ \bibnamefont{N\"agerl}}, \bibinfo {author}
  {\bibfnamefont{F.}~\bibnamefont{Ferlaino}}, \bibinfo {author}
  {\bibfnamefont{R.}~\bibnamefont{Grimm}}, \bibinfo {author}
  {\bibfnamefont{P.~S.}\ \bibnamefont{Julienne}},\ and\ \bibinfo {author}
  {\bibfnamefont{J.~M.}\ \bibnamefont{Hutson}},\ }%
  \bibfield{journal}{%
  \Doi{10.1103/PhysRevA.87.032517}{\bibinfo {journal} {Phys. Rev. A}}\ }%
  \textbf{\bibinfo {volume} {87}},\ \bibinfo {pages} {032517} (\bibinfo {month}
  {Mar}\ \bibinfo {year} {2013})%
  \bibAnnoteFile{NoStop}{Berninger2013fsf}%
\bibitem{Efimov1979lep}%
  \BibitemOpen
  \bibfield{author}{%
  \bibinfo {author} {\bibfnamefont{V.}~\bibnamefont{Efimov}},\ }%
  \bibfield{journal}{%
  \bibinfo {journal} {Sov. J. Nuc. Phys.}\ }%
  \textbf{\bibinfo {volume} {29}},\ \bibinfo {pages} {546} (\bibinfo {year}
  {1979})%
  \bibAnnoteFile{NoStop}{Efimov1979lep}%
\bibitem{Braaten2008tbr}%
  \BibitemOpen
  \bibfield{author}{%
  \bibinfo {author} {\bibfnamefont{E.}~\bibnamefont{Braaten}}, \bibinfo
  {author} {\bibfnamefont{H.-W.}\ \bibnamefont{Hammer}}, \bibinfo {author}
  {\bibfnamefont{D.}~\bibnamefont{Kang}},\ and\ \bibinfo {author}
  {\bibfnamefont{L.}~\bibnamefont{Platter}},\ }%
  \bibfield{journal}{%
  \Doi{10.1103/PhysRevA.78.043605}{\bibinfo {journal} {Phys. Rev. A}}\ }%
  \textbf{\bibinfo {volume} {78}},\ \bibinfo {pages} {043605} (\bibinfo {year}
  {2008})%
  \bibAnnoteFile{NoStop}{Braaten2008tbr}%
\bibitem{Berninger2011PhD}%
  \BibitemOpen
  \bibfield{author}{%
  \bibinfo {author} {\bibfnamefont{M.}~\bibnamefont{Berninger}},\ }%
  \emph{\bibinfo {title} {Universal three- and four-body phenomena in an
  ultracold gas of cesium atoms}},\ Ph.D. thesis,\ \bibinfo {school}
  {University of Innsbruck} (\bibinfo {year} {2011})%
  \bibAnnoteFile{NoStop}{Berninger2011PhD}%
\bibitem{SM}%
  \BibitemOpen
  \bibinfo {note} {See Supplemental Material for details on the sample
  preparation, the density and temperature determination, fitting of the decay
  curves, for additional information of the two sets of results, and for an
  updated fit of the $L_3$ data obtained previously \cite{Berninger2011uot} for
  the first Efimov resonance.}%
  \bibAnnoteFile{Stop}{SM}%
\bibitem{B_range}%
  \BibitemOpen
  \bibinfo {note} {The magnetic field range between 787 and 818 G corresponds
  to a range of scattering lengths between $-10^5$ and $-2600a_0$.}%
  \bibAnnoteFile{Stop}{B_range}%
\bibitem{Weber2003tbr}%
  \BibitemOpen
  \bibfield{author}{%
  \bibinfo {author} {\bibfnamefont{T.}~\bibnamefont{Weber}}, \bibinfo {author}
  {\bibfnamefont{J.}~\bibnamefont{Herbig}}, \bibinfo {author}
  {\bibfnamefont{M.}~\bibnamefont{Mark}}, \bibinfo {author}
  {\bibfnamefont{H.-C.}\ \bibnamefont{N\"agerl}},\ and\ \bibinfo {author}
  {\bibfnamefont{R.}~\bibnamefont{Grimm}},\ }%
  \bibfield{journal}{%
  \Doi{10.1103/PhysRevLett.91.123201}{\bibinfo {journal} {Phys. Rev. Lett.}}\
  }%
  \textbf{\bibinfo {volume} {91}},\ \bibinfo {pages} {123201} (\bibinfo {year}
  {2003})%
  \bibAnnoteFile{NoStop}{Weber2003tbr}%
\bibitem{Vonstecher2009sou}%
  \BibitemOpen
  \bibfield{author}{%
  \bibinfo {author} {\bibfnamefont{J.}~\bibnamefont{{von Stecher}}}, \bibinfo
  {author} {\bibfnamefont{J.~P.}\ \bibnamefont{D'Incao}},\ and\ \bibinfo
  {author} {\bibfnamefont{C.~H.}\ \bibnamefont{Greene}},\ }%
  \bibfield{journal}{%
  \Doi{10.1038/nphys1253}{\bibinfo {journal} {Nature Phys.}}\ }%
  \textbf{\bibinfo {volume} {5}},\ \bibinfo {pages} {417} (\bibinfo {year}
  {2009})%
  \bibAnnoteFile{NoStop}{Vonstecher2009sou}%
\bibitem{Ferlaino2009efu}%
  \BibitemOpen
  \bibfield{author}{%
  \bibinfo {author} {\bibfnamefont{F.}~\bibnamefont{Ferlaino}}, \bibinfo
  {author} {\bibfnamefont{S.}~\bibnamefont{Knoop}}, \bibinfo {author}
  {\bibfnamefont{M.}~\bibnamefont{Berninger}}, \bibinfo {author}
  {\bibfnamefont{W.}~\bibnamefont{Harm}}, \bibinfo {author}
  {\bibfnamefont{J.~P.}\ \bibnamefont{{D'Incao}}}, \bibinfo {author}
  {\bibfnamefont{H.-C.}\ \bibnamefont{N\"agerl}},\ and\ \bibinfo {author}
  {\bibfnamefont{R.}~\bibnamefont{Grimm}},\ }%
  \bibfield{journal}{%
  \Doi{10.1103/PhysRevLett.102.140401}{\bibinfo {journal} {Phys. Rev. Lett.}}\
  }%
  \textbf{\bibinfo {volume} {102}},\ \bibinfo {pages} {140401} (\bibinfo {year}
  {2009})%
  \bibAnnoteFile{NoStop}{Ferlaino2009efu}%
\bibitem{Zenesini2013rfb}%
  \BibitemOpen
  \bibfield{author}{%
  \bibinfo {author} {\bibfnamefont{A.}~\bibnamefont{Zenesini}}, \bibinfo
  {author} {\bibfnamefont{B.}~\bibnamefont{Huang}}, \bibinfo {author}
  {\bibfnamefont{M.}~\bibnamefont{Berninger}}, \bibinfo {author}
  {\bibfnamefont{S.}~\bibnamefont{Besler}}, \bibinfo {author}
  {\bibfnamefont{H.-C.}\ \bibnamefont{N\"{a}gerl}}, \bibinfo {author}
  {\bibfnamefont{F.}~\bibnamefont{Ferlaino}}, \bibinfo {author}
  {\bibfnamefont{R.}~\bibnamefont{Grimm}}, \bibinfo {author}
  {\bibfnamefont{C.~H.}\ \bibnamefont{Greene}},\ and\ \bibinfo {author}
  {\bibfnamefont{J.}~\bibnamefont{von Stecher}},\ }%
  \bibfield{journal}{%
  \Doi{10.1088/1367-2630/15/4/043040}{\bibinfo {journal} {New J. Phys.}}\ }%
  \textbf{\bibinfo {volume} {15}},\ \bibinfo {pages} {043040} (\bibinfo {year}
  {2013})%
  \bibAnnoteFile{NoStop}{Zenesini2013rfb}%
\bibitem{Fletcher2013soa}%
  \BibitemOpen
  \bibfield{author}{%
  \bibinfo {author} {\bibfnamefont{R.~J.}\ \bibnamefont{Fletcher}}, \bibinfo
  {author} {\bibfnamefont{A.~L.}\ \bibnamefont{Gaunt}}, \bibinfo {author}
  {\bibfnamefont{N.}~\bibnamefont{Navon}}, \bibinfo {author}
  {\bibfnamefont{R.~P.}\ \bibnamefont{Smith}},\ and\ \bibinfo {author}
  {\bibfnamefont{Z.}~\bibnamefont{Hadzibabic}},\ }%
  \bibfield{journal}{%
  \Doi{10.1103/PhysRevLett.111.125303}{\bibinfo {journal} {Phys. Rev. Lett.}}\
  }%
  \textbf{\bibinfo {volume} {111}},\ \bibinfo {pages} {125303} (\bibinfo
  {month} {Sep}\ \bibinfo {year} {2013})%
  \bibAnnoteFile{NoStop}{Fletcher2013soa}%
\bibitem{2nd_setB}%
  \BibitemOpen
  \bibinfo {note} {We use only set B for this purpose, because it contains more
  data at different hold times and therefore the fit converges better when both
  $L_\alpha$ and $\alpha$ are free in the fitting procedure.}%
  \bibAnnoteFile{Stop}{2nd_setB}%
\bibitem{alpha_heat}%
  \BibitemOpen
  \bibinfo {note} {The small deviation from $\alpha$ = 3 in region II can be
  explained by the heating effect, which leads to somewhat faster decay in the
  initial stage and a somewhat slower decay at the end of the hold time. This
  mimics higher-order loss.}%
  \bibAnnoteFile{Stop}{alpha_heat}%
\bibitem{Albritton1976ait}%
  \BibitemOpen
  \bibfield{author}{%
  \bibinfo {author} {\bibfnamefont{D.~L.}\ \bibnamefont{Albritton}}, \bibinfo
  {author} {\bibfnamefont{A.~L.}\ \bibnamefont{Schmeltekopf}},\ and\ \bibinfo
  {author} {\bibfnamefont{R.}~\bibnamefont{Zare}},\ }%
  in\ \emph{\bibinfo {booktitle} {Molecular Spectroscopy: Modern Research, Vol.
  II}},\ \bibinfo {editor} {edited by\ \bibinfo {editor} {\bibfnamefont{K.~N.}\
  \bibnamefont{Rao}}}\ (\bibinfo {publisher} {Academic Press},\ \bibinfo {year}
  {1976})%
  \bibAnnoteFile{NoStop}{Albritton1976ait}%
\bibitem{DIncao2009tsr}%
  \BibitemOpen
  \bibfield{author}{%
  \bibinfo {author} {\bibfnamefont{J.~P.}\ \bibnamefont{D'Incao}}, \bibinfo
  {author} {\bibfnamefont{C.~H.}\ \bibnamefont{Greene}},\ and\ \bibinfo
  {author} {\bibfnamefont{B.~D.}\ \bibnamefont{Esry}},\ }%
  \bibfield{journal}{%
  \Doi{10.1088/0953-4075/42/4/044016}{\bibinfo {journal} {J. Phys. B}}\ }%
  \textbf{\bibinfo {volume} {42}},\ \bibinfo {pages} {044016} (\bibinfo {year}
  {2009})%
  \bibAnnoteFile{NoStop}{DIncao2009tsr}%
\bibitem{Platter2009rct}%
  \BibitemOpen
  \bibfield{author}{%
  \bibinfo {author} {\bibfnamefont{L.}~\bibnamefont{Platter}}, \bibinfo
  {author} {\bibfnamefont{C.}~\bibnamefont{Ji}},\ and\ \bibinfo {author}
  {\bibfnamefont{D.~R.}\ \bibnamefont{Phillips}},\ }%
  \bibfield{journal}{%
  \Doi{10.1103/PhysRevA.79.022702}{\bibinfo {journal} {Phys. Rev. A}}\ }%
  \textbf{\bibinfo {volume} {79}},\ \bibinfo {pages} {022702} (\bibinfo {year}
  {2009})%
  \bibAnnoteFile{NoStop}{Platter2009rct}%
\bibitem{Thogersen2008upo}%
  \BibitemOpen
  \bibfield{author}{%
  \bibinfo {author} {\bibfnamefont{M.}~\bibnamefont{Th{\o}gersen}}, \bibinfo
  {author} {\bibfnamefont{D.~V.}\ \bibnamefont{Fedorov}},\ and\ \bibinfo
  {author} {\bibfnamefont{A.~S.}\ \bibnamefont{Jensen}},\ }%
  \bibfield{journal}{%
  \Doi{10.1103/PhysRevA.78.020501}{\bibinfo {journal} {Phys. Rev. A}}\ }%
  \textbf{\bibinfo {volume} {78}},\ \bibinfo {pages} {020501(R)} (\bibinfo
  {year} {2008})%
  \bibAnnoteFile{NoStop}{Thogersen2008upo}%
\bibitem{Wang2014uvd}%
  \BibitemOpen
  \bibfield{author}{%
  \bibinfo {author} {\bibfnamefont{Y.}~\bibnamefont{Wang}}\ and\ \bibinfo
  {author} {\bibfnamefont{P.~S.}\ \bibnamefont{Julienne}},\ }%
  \bibfield{journal}{%
  \bibinfo {journal} {arXiv:1404.0483}}%
   (\bibinfo {year} {2014})%
  \bibAnnoteFile{NoStop}{Wang2014uvd}%
\bibitem{wangjulienne}%
  \BibitemOpen
  \bibfield{author}{%
  \bibinfo {author} {\bibfnamefont{Y.}~\bibnamefont{Wang}}\ and\ \bibinfo
  {author} {\bibfnamefont{P.~S.}\ \bibnamefont{Julienne}},\ }%
  \bibinfo {note} {private communication}%
  \bibAnnoteFile{NoStop}{wangjulienne}%
\bibitem{Jonsell2002ubo}%
  \BibitemOpen
  \bibfield{author}{%
  \bibinfo {author} {\bibfnamefont{S.}~\bibnamefont{Jonsell}}, \bibinfo
  {author} {\bibfnamefont{H.}~\bibnamefont{Heiselberg}},\ and\ \bibinfo
  {author} {\bibfnamefont{C.~J.}\ \bibnamefont{Pethick}},\ }%
  \bibfield{journal}{%
  \Doi{10.1103/PhysRevLett.89.250401}{\bibinfo {journal} {Phys. Rev. Lett.}}\
  }%
  \textbf{\bibinfo {volume} {89}},\ \bibinfo {pages} {250401} (\bibinfo {month}
  {Nov}\ \bibinfo {year} {2002})%
  \bibAnnoteFile{NoStop}{Jonsell2002ubo}%
\bibitem{Levinsen2014etu}%
  \BibitemOpen
  \bibfield{author}{%
  \bibinfo {author} {\bibfnamefont{J.}~\bibnamefont{Levinsen}}, \bibinfo
  {author} {\bibfnamefont{P.}~\bibnamefont{Massignan}},\ and\ \bibinfo {author}
  {\bibfnamefont{M.~M.}\ \bibnamefont{Parish}},\ }%
  \bibfield{journal}{%
  \bibinfo {journal} {arXiv:1402.1859}}%
   (\bibinfo {year} {2014})%
  \bibAnnoteFile{NoStop}{Levinsen2014etu}%
\bibitem{Zenesini2014vdw}%
  \BibitemOpen
  \bibfield{author}{%
  \bibinfo {author} {\bibfnamefont{A.}~\bibnamefont{Zenesini}}, \bibinfo
  {author} {\bibfnamefont{B.}~\bibnamefont{Huang}}, \bibinfo {author}
  {\bibfnamefont{M.}~\bibnamefont{Berninger}}, \bibinfo {author}
  {\bibfnamefont{H.-C.}\ \bibnamefont{N\"{a}gerl}}, \bibinfo {author}
  {\bibfnamefont{F.}~\bibnamefont{Ferlaino}},\ and\ \bibinfo {author}
  {\bibfnamefont{R.}~\bibnamefont{Grimm}},\ }%
  \bibinfo {note} {to be published.}%
  \bibAnnoteFile{Stop}{Zenesini2014vdw}%
\bibitem{Dincao2006eto}%
  \BibitemOpen
  \bibfield{author}{%
  \bibinfo {author} {\bibfnamefont{J.~P.}\ \bibnamefont{D'Incao}}\ and\
  \bibinfo {author} {\bibfnamefont{B.~D.}\ \bibnamefont{Esry}},\ }%
  \bibfield{journal}{%
  \Doi{10.1103/PhysRevA.73.030703}{\bibinfo {journal} {Phys. Rev. A}}\ }%
  \textbf{\bibinfo {volume} {73}},\ \bibinfo {pages} {030703(R)} (\bibinfo
  {year} {2006})%
  \bibAnnoteFile{NoStop}{Dincao2006eto}%
\bibitem{Repp2013ooi}%
  \BibitemOpen
  \bibfield{author}{%
  \bibinfo {author} {\bibfnamefont{M.}~\bibnamefont{Repp}}, \bibinfo {author}
  {\bibfnamefont{R.}~\bibnamefont{Pires}}, \bibinfo {author}
  {\bibfnamefont{J.}~\bibnamefont{Ulmanis}}, \bibinfo {author}
  {\bibfnamefont{R.}~\bibnamefont{Heck}}, \bibinfo {author}
  {\bibfnamefont{E.~D.}\ \bibnamefont{Kuhnle}}, \bibinfo {author}
  {\bibfnamefont{M.}~\bibnamefont{Weidem\"uller}},\ and\ \bibinfo {author}
  {\bibfnamefont{E.}~\bibnamefont{Tiemann}},\ }%
  \bibfield{journal}{%
  \Doi{10.1103/PhysRevA.87.010701}{\bibinfo {journal} {Phys. Rev. A}}\ }%
  \textbf{\bibinfo {volume} {87}},\ \bibinfo {pages} {010701} (\bibinfo {month}
  {Jan}\ \bibinfo {year} {2013})%
  \bibAnnoteFile{NoStop}{Repp2013ooi}%
\bibitem{Tung2013umo}%
  \BibitemOpen
  \bibfield{author}{%
  \bibinfo {author} {\bibfnamefont{S.-K.}\ \bibnamefont{Tung}}, \bibinfo
  {author} {\bibfnamefont{C.}~\bibnamefont{Parker}}, \bibinfo {author}
  {\bibfnamefont{J.}~\bibnamefont{Johansen}}, \bibinfo {author}
  {\bibfnamefont{C.}~\bibnamefont{Chin}}, \bibinfo {author}
  {\bibfnamefont{Y.}~\bibnamefont{Wang}},\ and\ \bibinfo {author}
  {\bibfnamefont{P.~S.}\ \bibnamefont{Julienne}},\ }%
  \bibfield{journal}{%
  \Doi{10.1103/PhysRevA.87.010702}{\bibinfo {journal} {Phys. Rev. A}}\ }%
  \textbf{\bibinfo {volume} {87}},\ \bibinfo {pages} {010702} (\bibinfo {month}
  {Jan}\ \bibinfo {year} {2013})%
  \bibAnnoteFile{NoStop}{Tung2013umo}%
\bibitem{Tung2014oog}%
  \BibitemOpen
  \bibfield{author}{%
  \bibinfo {author} {\bibfnamefont{S.-K.}\ \bibnamefont{Tung}}, \bibinfo
  {author} {\bibfnamefont{K.}~\bibnamefont{Jimenez-Garcia}}, \bibinfo {author}
  {\bibfnamefont{J.}~\bibnamefont{Johansen}}, \bibinfo {author}
  {\bibfnamefont{C.~V.}\ \bibnamefont{Parker}},\ and\ \bibinfo {author}
  {\bibfnamefont{C.}~\bibnamefont{Chin}},\ }%
  \bibfield{journal}{%
  \bibinfo {journal} {arXiv:1402.5943}}%
   (\bibinfo {year} {2014})%
  \bibAnnoteFile{NoStop}{Tung2014oog}%
\bibitem{Pires2014ooe}%
  \BibitemOpen
  \bibfield{author}{%
  \bibinfo {author} {\bibfnamefont{R.}~\bibnamefont{Pires}}, \bibinfo {author}
  {\bibfnamefont{J.}~\bibnamefont{Ulmanis}}, \bibinfo {author}
  {\bibfnamefont{S.}~\bibnamefont{H\"{a}fner}}, \bibinfo {author}
  {\bibfnamefont{M.}~\bibnamefont{Repp}}, \bibinfo {author}
  {\bibfnamefont{A.}~\bibnamefont{Arias}}, \bibinfo {author}
  {\bibfnamefont{E.~D.}\ \bibnamefont{Kuhnle}},\ and\ \bibinfo {author}
  {\bibfnamefont{M.}~\bibnamefont{Weidem\"{u}ller}},\ }%
  \bibfield{journal}{%
  \bibinfo {journal} {arXiv:1403.7246}}%
   (\bibinfo {year} {2014})%
  \bibAnnoteFile{NoStop}{Pires2014ooe}%
\end{thebibliography}

%

\end{document}